\documentclass [11pt] {article}
\usepackage{amssymb}
\usepackage{latexsym}
\usepackage{amsmath, bbm, bm}

\title{The Algebra of Observables of the closed Bosonic String \\ in 
$(1+3)$-dimensional Minkowski-Space: \\updating the structural Analysis}

\author{
\null\\[-4mm]
{\bf K. Pohlmeyer}
\\[4mm]
Fakult\"at f\"ur Mathematik und Physik,
Physikalisches Institut,
Universit\"at Freiburg, \\
Hermann-Herder-Stra{\ss}e 3,
D-79104 Freiburg, Germany
\\[3mm]
}

\date{March 2006}

\newcommand {\cP}  {\mathcal P}

\newcommand{\fa}{\mathfrak a}
\newcommand{\A}{\mathcal A}
\newcommand{\G}{G}
\newcommand{\g}{\mathfrak g}
\newcommand{\bG}{\G}
\newcommand{\h}{\mathfrak h}
\newcommand{\R}{\mathcal R}
\newcommand{\U}{\mathfrak U}
\newcommand{\Z}{\mathcal Z}
\newcommand{\grem}{\G^{\mathrm{rem}}}
\newcommand{\fr}{\mathfrak R}
\newcommand{\fZ}{\mathfrak Z}

\newcommand{\Zn}[1]{\mathcal Z^{(#1)}}

\newcommand{\N}{\mathbbm N}
\newcommand{\ZN}{\mathbbm Z}
\newcommand{\RN}{\mathbbm R}

\newcommand{\qI}[3]{\hat{\mathcal{#1}}_{#2}^{#3} \hspace{-3.7ex}\backslash \hspace{2.2ex}}
\newcommand{\cI}[3]{\mathcal{#1}_{#2}^{#3} }

\newcommand{\Rt}{\mathcal R^t}

\newcommand{\V}{\mathfrak h}

\newcommand{\shuff}[3]{#1_{\small\frame{${\scriptscriptstyle
\begin{array}{l} #2 \\ #3 \end{array}}$}}}

\newcommand{\shuffzB}[5]{#1_{\mbox{\small\frame{${\scriptscriptstyle
\begin{array}{l} #2 \\ #3 \end{array}}$}}
\ \mbox{\small\frame{${\scriptscriptstyle
\begin{array}{l} #4 \\ #5 \end{array}}$}}}}

\begin{document}

\maketitle

\begin{abstract}
The purpose of the present paper is the communication of some results and observations which shed new light on the algebraic structure of the algebra of string observables both in the classical and in the quantum theory.
\end{abstract}


A suitable introduction to the construction and investigation of the algebra of observables of the Nambu-Goto string in $D$-dimensional Minkowski-Space can be found in \cite{1} and \cite{2}. Reference \cite{1} is the starting point for the present communication which will deal only with the case dimension $D=1+3$. Moreover, the mass-square $m^2=\cP^{\mu}\cP_{\mu}$ will be assumed to be positive and the rest-frame $\cP_{\mu}=m\;\delta_{\mu,0}$ will be chosen as the reference system. Here, the symbols $\cP_{\mu}$, $\mu \in \{0,1,2,3\}$ denote the components of the energy-momentum vector of the string.

We use the same concepts, methods, definitions and notations as the authors of \cite{1}. In particular, the building units of the present analysis are the monodromy variables $\Rt_{\mu_1\ldots \mu_M}$ with $\mu_i \in \{0,1,2,3\}, \; i \in \{1,\ldots, M\}, \; M \in \mathbbm N, \; M\geq 1$, and we identify $\Rt_{\mu}$ with the components $\cP_{\mu}$ of the energy-momentum vector: $\Rt_{\mu}=\cP_{\mu}$. In the following, the objects $\Rt_{\mu_1\ldots \mu_M}$ are called truncated tensors ({\it recte:} tensor components) of tensor degree $M$. The only algebraic relations between them are linear:
\[
\shuff{\Rt}{\mu_1\ldots \mu_L}{\mu_{L+1}\ldots \mu_M} =0,\quad 0<L<M, \quad M\geq 2,
\]
where {\small{\frame{${\scriptscriptstyle
\begin{array}{l} {\mu_1\ldots \mu_L}\\{\mu_{L+1}\ldots \mu_M}\end{array}}$}}} denotes the sum over the shuffle permutations of the \lq deck{s\rq} of indices $\mu_1\ldots\mu_L$ and $\mu_{L+1}\ldots\mu_M$, respectively. These relations can be completely resolved by considering the linear span $\fr_{(M)}$ of all truncated tensors of fixed tensor rank $M$ -- this linear space has dimension $n(4,M)$ -- and by introducing a suitable basis: $\big\{\Rt_{\mu_1\ldots\mu_M} \ \big| $ the sequences $\mu_1\ldots\mu_M$ being cyclically minimal when the numerical values for the indices $\mu_i$ are inserted$\big\}$. The symbol $n(4,M)$ stands for 
\[
n(4,M)=\frac{1}{M}\sum_{D{\bf|}M} {\mu}(D) 4^{\frac{M}{D}},
\]
where the sum extends over all divisors $D$ of $M$ and where $\mu$ denotes the M\"obius function.
Consider the direct sum $\bm{\fr}:= \bigoplus_{M=1}^\infty \fr_{(M)}$. The modified Poisson bracket $\{\;,\;\}_*$,
\begin{eqnarray*}
\{\Rt_{\mu_1\dots\mu_M}, \Rt_{\nu_1\dots\nu_N}\}_* &=& \sum_{i=1}^M\sum_{j=1}^N \; 2\; g_{\mu_i\nu_j}(-1)^{M+i+j}
\shuffzB{\Rt}{\mu_1\dots\mu_{i-1}}
{\mu_M\dots\mu_{i+1}}{\nu_{j-1}\dots\nu_1}{\nu_{j+1}\dots\nu_N}\ \quad  \forall M,N \geq 2\\
\mbox{and} \quad \{\Rt_\mu,\Rt_{\nu_1\dots\nu_N}\}_*&=&0 \quad \forall N \in \N
\end{eqnarray*}
endows $\fr$ with the structure of a graded Lie algebra:
\[
\fr \times\fr\to\fr: \quad \fr_{(l_1+2)}\times\fr_{(l_2+2)}\to \fr_{(l_1+l_2+2)},\quad l_k \in \{-1,0,1,\ldots\}, k=1,2\ .
\]
Before we shall report more detailed properties of the Lie algebra $\fr$, we turn to one of its Lie subalgebras: the subalgebra of so-called space-like truncated tensors
\[
\mathcal{LH}\left\{\Rt_{\mu_1\dots\mu_M}\ \big|\ \mu_k \in \{1,2,3\}\; \mbox{for} \; k \in \{1,M\},\; \mu_k \in\{0,1,2,3\} \; \mbox{for}\;k \in\{2,\ldots,M-1\},\; M\in \N\right\}.
\]
This subalgebra is of paramount importance for the construction and the analysis of the algebra of string observables.

We obtain a basis of this subalgebra in two steps: first, we consider the
linear span of the space-like truncated tensors for each tensor rank
$M$ separately. We resolve the linear relations between the space-like
truncated tensors in each such linear span (i.e. the only algebraic
relations among the space-like truncated tensors existing at all), by
introducing a suitable basis: for the linear subspan of those
truncated tensors of tensor rank M for which {\em none} of the indices
$\{\mu_1,\dots,\mu_M\}$ takes on the value zero, we proceed as before
and obtain $n(3,M)$ basis elements. For the linear subspan of the
remaining truncated tensors where {\em at least one} of the indices
$\{\mu_2,\ldots,\mu_{M-1}\}$ takes on the value zero, we obtain
$n(4,M)-n(4,M-1) - n(3,M)$ basis elements by assigning exactly one
truncated tensor $\Rt_{\mu_1\dots\mu_{j-1}0\mu_{j+1}\dots\mu_M}$ to
every cyclically minimal but \underline{not} doubly cyclically minimal sequence
$0\nu_2\ldots\nu_M$ such that
\begin{eqnarray*}
0\,
\mbox{max (all values obtained from the decks }
 \mu_{j+1}\ldots \mu_M 
\mbox{ and } \mu_{j-1}\ldots \mu_1 \mbox{ by a shuffle permutation)}
\\ \quad = \  0\, \nu_2\ldots\nu_M	\ .
\end{eqnarray*}

Here, by definition, a sequence $\mu_1\ldots\mu_M$ is called \underline{doubly cyclically minimal} if both the sequence $\mu_1\ldots\mu_M$ and the sequence $\mu_2\ldots\mu_M$ are cyclically minimal.
\vskip0.5cm

We then consider the direct sum of the former and the latter linear subspan and hence obtain the desired linear span $\G_{(M)}$ of dimension $n(4,M)-n(4,M-1)$ for $M>1$. We define $\G_{(1)}$ als the linear span of $\R^t_\mu=\cP_\mu$. Now we form the direct sum $\bm{\G}:=\bigoplus_{M=1}^{\infty} \G_{(M)}$. The modified Poisson bracket $\{\;,\;\}_*$ endows $\bG$ with the structure of a graded Lie algebra, embedded in the Lie algebra $\fr$:
\[
\bG\times\bG\to \bG:\quad  \G_{(l_1+2)}\times \G_{(l_2+2)}\to\G_{(l_1+l_2+2)}.
\]
$\G_{(l+2)}$ is the germ of the stratum $\V^l$ of the Poisson algebra (and the quantum algebra) of observables, $l$ corresponding to the degree of the gradation (and the filtration, respectively).

The following facts are known about the algebra $\bG$:
\begin{enumerate}
\item[1.)] The direct summands $\G_{(l+2)}$, $l=-1,0,+1,+2,\ldots$, are representation spaces of $\mathrm O(3)$. In fact, the Lie algebra of the infinitesimal generators of $\mathrm O(3)$ can be identified with $\G_{(2)}$.
\item[2.)] $\bG$ is {\em not} finitely generated. It is well known that in every direct summand $\G_{(l+2)}$ where $l$ is an odd natural number, there is {\em at least} one element, for instance: the so-called exceptional element $\sum_{j=1}^3\R_{j \! \underbrace{\scriptstyle 0\dots0}_{l \ {\rm times}}\! j}$, $l=$~odd natural number, which cannot be generated by the (modified) Poisson bracket operation from the elements of the linear spaces $\G_{(l'+2)}$ with $l'<l$.
\end{enumerate}

It has been conjectured repeatedly that the set of basis elements of
$\G_{(l+2)}, \; l \in \{-1,0,+1\}$ augmented by the set of all
exceptional elements of $\G$ provides a complete system of generators
of the algebra $\G$. Though this conjecture - in a suitably modified
form - is correct for the direct sum of the spaces $\G_{(l'+2)}$ with
$l'\le8$, it is wrong for the remainder of $\G$. For the case $l=9$,
the previously mentioned augmented set of generators produces only a
subspace of $\G_{(9+2)}$ with a dimension falling short of the
dimension of $\G_{(9+2)}$ by one. One more generator for $\G_{(9+2)}$
has to be added to the list, for instance:
\[
\frac{1}{2}\sum_{j=1}^3\sum_{k=1}^3\Rt_{\begin{array}{c}
\frame{\tiny ${\scriptscriptstyle \begin{array}{c} j 0 \\ k 0 \end{array}}$}\end{array} 0 0 0 \begin{array}{c}
\frame{\tiny ${\scriptscriptstyle \begin{array}{c} 0 j \\ 0 k \end{array}}$}\end{array} }.
\]
Probably, this example marks only the beginning of a new infinite
series of generators of the algebra $\G$, possibly the next member of
this series being contained in the direct summand
$\G_{(15+2)}$. Moreover, one must face the unpleasant scenario that
the generation of $\G$ may require an infinite number of infinite
series of generators. However, this structure of the Lie algebra $\G$
does not necessarily entail a calamity for the quantum algebra of
observables, which is only filtered, in sharp contrast to the
classical Poisson algebra of observables, which is graded, and where,
indeed, it does create enormous difficulties (see below).

Please note that the generators identified so far beyond those
contained in $\bigoplus_{l=-1,0,+1}\G_{(l+2)}$ are invariant under the
action of the group $\mathrm O(3)$ . This is not merely a
coincidence. It can be proved that {\em all} generators of $\G$ beyond
those which are contained in $\bigoplus_{l=-1,0,+1}\G_{(l+2)}$ may be
chosen as $\mathrm O(3)$ scalars. The proof of this and other related
statements made in the present communication will be submitted for
publication in an appropriate mathematical-physics journal in due
course.

It is not difficult to prove that the Lie algebra $\G$ can be
decomposed into a semidirect sum of an Abelian algebra $\A$ generated
by the exceptional elements $\sum_{j=1}^3 \Rt_{j \! 
\underbrace{\scriptstyle 0\dots0}_{l \ {\rm times}}\! j}$,
$l\in\{1,3,5,\ldots\}$, and the Lie algebra $\G^{\mathrm{rem}}$
generated by the remaining generators of $\G$:
$\G^{\mathrm{rem}}=\bigoplus_{l=-1}^{\infty}\G^{\mathrm{rem}}_{(l+2)}$
(the former algebra acting on the latter one).  As for a description
of the Lie algebra $\G^{\mathrm{rem}}$ in terms of generators and
generating relations, the situation is as follows: the action of all
the exceptional elements on the generators $\G^{\mathrm{rem}}$ which
are contained in the direct summands of
$\bigoplus_{l=-1,0,+1}\G^{\mathrm{rem}}_{(l+2)}$ is explicitly known
and so are all the {\em independent} generating relations of
$\bigoplus_{l=-1}^6\G^{\mathrm{rem}}_{(l+2)}$ (independent if the
assistance of the elements of $\A$ is declined) and all the {\em truly
independent} generating relations of
$\bigoplus_{l=-1}^6\G^{\mathrm{rem}}_{(l+2)}$ (truly independent if
the assistance of the elements of $\A$ is included).

It is \underline{conjectured}, that, in addition to the relations
specifying the actions of the exceptional elements on the infinitely
many new generators and the actions of the new generators on each
other and on the generators contained in
$\bigoplus_{l=-1,0,+1}\G^{\mathrm{rem}}_{(l+2)}$ , there will arise
infinitely many new truly independent generating $J^P=0^+$-relations
of $\G^{\mathrm{rem}}$ correlated to the emergence of the new
generators, in particular to their respective degrees. The rationale
behind this conjecture will be explained below.
\vskip0.5cm

Now, returning to the embedding graded Lie algebra $\bm{\fr}:= \bigoplus_{l=-1}^\infty \fr_{(l+2)}$, we notice that also each of its direct summands $\fr_{(l+2)}$ carries a linear representation of the group $\mathrm O(3)$. Thus each $\fr_{(l+2)}$ can be decomposed into isotypical $\mathrm O(3)$ components. Making extensive use of this $\mathrm O(3)$-covariance and of the information obtained by the analysis of the structure of the embedded Lie algebra $\grem$ up to the year 2003, it has been checked explicitly that$\bigoplus_{l=-1}^5\fr_{(l+2)}$ is generated by the elements $\Rt_{\mu}=\cP_{\mu},\; \mu=0,1,2,3$ of $\fr_{(-1+2)}$, $\Rt_{0j}, \; j=1,2,3$ and $\Rt_{jk}, \; j,k=1,2,3$ of $\fr_{(0+2)}$ and the elements of $\fr_{(1+2)}$ with spin magnetic quantum number zero.

The check - though  amounting to a couple of hundred separate computations - did not require much skill. In fact, the vast majority  of the separate computations could be carried out with ease in one line.

Also, by the same techniques which were applied before to the Lie algebra $\G$, it can be proved that the only potential candidates for additional generators of $\fr$ must be $\mathrm O(3)$-invariant elements.

Judging by this experience, it seems likely to me that the entire Lie algebra $\fr$ is finitely generated by the generators specified above. However, in sharp contrast  to the ease of the explicit check mentioned above, a general proof of this speculation may require considerable combinatorical skill. A confirmation of this speculation by a proof would establish a possibly helpful connection between the structural analysis of the algebra of string observables and Hilbert's fourteenth problem~\cite{3}.
\vskip0.5cm

Next, we pass to the symmetric enveloping algebra without unit $\mathrm S(\bm{\fr})$, which forms a Poisson algebra. Actually, we are interested in the Poisson subalgebra $\h$ of $\mathrm S(\bm{\fr})$ produced by those polynomials in the truncated tensors at the given fixed point $(\tau, \sigma)$, $\tau \in \RN^1$, $\sigma \in S^1$, which do {\em not} depend on $\tau$ and $\sigma$ any more.
These polynomials are called reparametrization invariants, or for short: invariants. The invariants are of the form
\[
\label{zdefinition1}
\Z_{\mu_1\ldots\mu_M}=\sum_{K=1}^M \Zn{K}_{\mu_1\ldots\mu_M} \qquad {\rm with}
\]
\[
\Zn{K}_{\mu_1\ldots\mu_M}=\frac{1}{K!} \; \fZ_M\circ\left(\sum_{\scriptstyle{1\leq a_1<\ldots<a_{K-1}<M}}\Rt_{\mu_1\ldots\mu_{a_1}}\; \Rt_{\mu_{a_1+1}\ldots\mu_{a_2}}\ldots\Rt_{\mu_{a_{K-1}+1}\ldots\mu_{M}}\right)
\]
where $\fZ_M$ denotes the sum over the cyclic permutations of the indices $\mu_1,\ldots,\mu_M$. The polynomials $\Zn{K}_{\mu_1\ldots\mu_M}$ are also invariants, called homogeneous invariants of tensor rank $M$ and degree of homogeneity $K$.

The term $\frac{1}{K!} \;\fZ_M\circ\left(\cP_{\mu_1}\cP_{\mu_2}\ldots\cP_{\mu_{K-1}}\Rt_{\mu_K\ldots\mu_M}\right)$ is called the leading part of $\Zn{K}_{\mu_1\ldots\mu_M}$. In the momentum rest frame, the homogeneous invariant $\Zn{K}_{\mu_1\ldots\mu_M}$ can be completely reconstructed
from just one non-vanishing term of the sum $\frac{1}{K!} \;\fZ_M\circ\left(\ldots\right)$ belonging to the leading part: only those terms in the sum $\frac{1}{K!} \;\fZ_M\circ\left(\ldots \right)$ survive for which $\mu_{[j]}=\mu_{[j+1]}=\ldots=\mu_{[j+k-2]}=0$ where the symbol $[k]$ denotes the number $k$ modulo $M$. The sum simplifies to

\begin{equation}
\label{simplifiedzsum}
\frac{m^{(K-1)}}{K!}\sum_{j:\mu_{[j]}=\ldots=\mu_{[j+k-2]}=0}\Rt_{\mu_{[j+k-1]}
\ \ldots\ \mu_{[M+j-1]}},
\end{equation}
the latter sum producing a non-trivial element of $\G_{(l+2)}$ with $l=M-K-1$ if the sequence $\mu_1\ldots \mu_M$ involves two or more non-zero valued indices, and zero otherwise, unless $M=1$.

Thus choosing a suitable natural number $K$ for each adequately normalized
basis element of $\G_{(l+2)}$, one can come up with a homogeneous
invariant of tensor rank $l+K+1$ with degree of homogeneity $K$ and
degree of gradation $l$, called standard invariant \cite{4}. The
linear span of the homogeneous invariants of degree $l$ shall be
denoted by $\V^l$, such that $\h=\bigoplus_{l=-1}^{\infty}\V^l$. Under
Poisson bracket operation $\h\times\h\to\h$, we have
\[
\V^{l_1}\times\V^{l_2}\to \V^{l_1+l_2}.
\]
Under tensor multiplication $\h\times\h\to\h$, we have
\[
\V^{l_1}\times\V^{l_2}\to \V^{l_1+l_2+1}.
\]
Thus a reasonably explicit basis (w.r.t. tensor multiplication) of the
graded Poisson algebra $\h$ of algebraically independent elements is
available in the form of the complete set of standard invariants.
\vskip0.5cm

As for the description of the Poisson algebra $\h$ in terms of
generators and generating relations, the situation corresponds closely
to the one of the Lie algebra $\G$: each generator of $\G$ brings up a
generator of $\h$ in its train and vice versa, and the same holds for
the the generating relations. The mapping of a given generator of $\G$
with degree $l$ to a generator of $\h$ is a lift which leaves some
freedom, namely the addition of \underline{nonlinear}
degree-$l$-monomials in the standard invariants.

There is still no general proof that the Poisson algebra $\h$
possesses the structure corresponding to the decomposition of the Lie
algebra $\G$ into a semi-direct sum of the abelian Lie algebra $\A$,
generated by the exceptional elements, and the Lie algebra $\grem$. In
fact, in view of the results of \cite{5} on the quantum algebra of
string observables, interest has somewhat weakened regarding the
existence a Lie algebra $\g$ with the following properties:
\begin{enumerate}
\item[1.)] $\h$ coincides with the symmetric algebra obtained from $\g$,
\item[2.)] $\g$ is the semi-direct sum of an Abelian algebra $\fa$ generated by suitably chosen counterparts of the generators of $\A$ (in other words, of the exceptional elements), and the Poisson algebra $\U$ generated by the rest of the generators of $\h$: $\g=\fa\ltimes\U$, and
\item[3.)] the generators of $\fa$ act as derivations on the Poisson algebra $\U$.
\end{enumerate}

The last comment on the classical algebras $\fr$ and $\h$ to be made
in this paper will concern a rigorous proof of {\bf Proposition 17} in
\cite{4}. Roughly, this proposition claims that in the energy-momentum
rest frame $\cP_{\mu}=m\;\delta_{\mu,0},\; m>0$, every polynomial
$\cP_{\mathrm{inv}} $ in the truncated tensors
$\Rt_{\mu_1\ldots\mu_M}$ which is invariant under reparamerizations,
is a polynomial $Q$ (possibly with $m$-dependent coefficients) in the
standard invariants $\Z_{k+1,i},\; 
i=1,\ldots,I_{K-1}$, $1 \leq I_{K-1} \leq n(4,K+1) - n(4,K)$, $K \in 
\mathbbm N$. Up to now, the `Indication of the proof' in \cite{4} has not been
replaced by a complete and conclusive presentation of the
argument. The proof announced here follows a different line of
argumentation strongly relying on two properties of the standard
invariants: their {\em algebraic} independence~\cite{4} and their
completeness~\cite{6}. Thus it fills a gap and guarantees that the
recursive algorithm for the determination of the polynomial $Q$ in the
standard invariants for any given polynomial $\cP_{\mathrm{inv}}$ does
not leave any remainder. The said proof also provides an efficient
non-recursive algorithm for the determination of $Q$. The
details of this proof can be found in the appendix of the present paper.
\vskip0.5cm

Now, let us turn to the construction and structural analysis of a
quantum algebra of string observables. In Ref.~\cite{7}, a recursive
deformation routine has been proposed for the passage from the
classical algebra of string observables, i.e. the Poisson algebra
$\h$, to an algebra of quantum string observables, envisaged as the
non-commutative enveloping Lie algebra $\Hat{\h}$ generated by the
quantum counterparts of the classical generators subject to the
quantum counterparts of the classical truly independent generating
relations. This deformation routine was organized in terms of
successive cycles of step by step increasing degrees $l>1$. The main
ingredients of the routine were the correspondence principle,
structural similarity between $\h$ and $\Hat{\h}$, and consistency (in
particular w.r.t. the Jacobi identity). The cycles carried out so far
are of degrees $2$ to $5$ corresponding to the powers $\hbar^4$ to
$\hbar^7$ in Planck's constant. For the symbolic computations and
their results, see \cite{7} concerning the degrees $2$ and $3$,
\cite{8} and \cite{5} concerning the degrees $4$ and $5$,
respectively.

The above investigations lead to the following revision of the original concepts over the past years:
\begin{enumerate}
\item[(a)] the routine is not likely to produce a unique quantum algebra of string observables. Instead, it is likely to produce a {\em family} of such quantum algebras, labelled by rational valued parameters: first one such parameter $f$ \cite{7}, then three such parameters $f,g_1$ and $g_2$, \cite{8}, and next possibly even more such parameters.
\item[(b)] the structural similarity requirement applied to the semi-direct splitting $\g=\fa\ltimes \U$ cannot be upheld \cite{5}.
\end{enumerate}

\underline{As to (a)}: the authors of \cite{8} set out to define the 
quantum generator $\qI{B}{0}{(3)}$ -- corresponding to the 
degree-3-generator $\cI{B}{0}{(3)}$ of $\fa$ -- as a derivation. For that 
purpose they considered the Poisson actions of $\cI{B}{0}{(3)}$ on the 
generators of $\h$ contained in $\V^1$. They added to them the most 
general quantum corrections compatible with the correspondence principle , 
i.e. sixty-one correction terms of degree $l$, $0\leq l\leq 4$ and of 
appropriate spins and parities. Then on the resulting candidates for the 
various quantum actions the following consistency requirements were 
imposed: \begin{enumerate} \item the vanishing of the quantum action of 
$\qI{B}{0}{(3)}$ on $\qI{B}{0}{(1)}$, and \item the Jacobi identity, which 
by necessity involves the only free paramter $f$ of the previous cycles of 
the construction routine of the algebra $\Hat{\h}$. \end{enumerate} This 
reduces the number of free parameters from 61 to 3: apart from $f$, two 
new parameters $g_1$ and $g_2$ remain undetermined.

The generating relations of $\h$ carrying a degree $l$ w.r.t. an $\N$-gradation correspond to generating relations of $\Hat{\h}$ with the same degree $l$, this time, however, w.r.t. an $\N$-filtration and w.r.t. a $\ZN_2$-gradation ($l=$ even or odd). Realizing this, it is only logical to allow for a similar kind of relationship between the generators of $\h$ carrying a degree w.r.t. an $\N$-gradation  and the corresponding generators of $\Hat{\h}$ carrying the same degree $l$, this time, however, w.r.t. an $\N$-filtration and w.r.t. a $\ZN_2$-gradation.

Exploiting the freedom which comes along with this point of view, on the
basis of the results of~\cite{8} we observe that the additional free
paramters $g_1$ and $g_2$ -- among other free parameters like $f$
allegedly labelling the quantum algebras of string observables -- can
be completely removed from $\Hat{\h}$, for instance by the following
redefinition
\[
\qI{B}0{(3)}_{\!\!\! \mbox{\scriptsize{\cite{8}}}}\  \longrightarrow 
\qI{B}0{(3)} := \qI{B}0{(3)}_{\!\!\! \mbox{\scriptsize{\cite{8}}}}\  - g_1 
\qI{B}0{(1)}_{\!\!\! \mbox{\scriptsize{\cite{7}}}}
-\frac 1 2\sqrt{3}\ g_2 \ 
    \{ \qI{J}{1}{\phantom{(0)}}\!\!\!\! , \qI{J}{1}{\phantom{(0)}}\!\!\!\!  \}_0 
\]
or by an equivalent redefinition (see below). Then for the time being, we are left with only one free parameter, $f$, labelling the members of the class of quantum algebras of string observables.
\vskip0.5cm

\underline{As to (b)}: in \cite{5} the cycle of degree 5 of the quantum deformation routine has been carried out. In particular, the only truly independent classical generating relation of degree 5, a $J^P=0^+$ relation, has been accordingly deformed. The result is deposited in the electronic archive [math-ph/0210024]. (According to the preceeding comment, without loss of generality the result may be slightly simplified by setting the parameters $g_1$ and $g_2$ equal to zero.) The result clearly shows that the quantum algebra of string observables does not even allow for a partition of its generators into two subsets: the quantum counterparts of the exceptional ones and of the non-exceptional ones, such that the following is true: restricting the filtration degrees to five or less, the monomials in the generators and the (multiple) commutators thereof --  (apart from the linear cases) written as (multiple) anticomutators -- formed exclusively for the first one of the two subsets are linearly independent of those formed exclusively for the second one. A fortiori, the original hypothesis motivated by an expected structural similarity between $\h$ and $\Hat{\h}$, must be given up. However, we should not miss the lesson we can draw from the quantum deformation of this $l=5$, $J^P=0^+$ classical relation: in striking contrast of the r\^ole of $\cI{B}0{(3)}$ in the graded Poisson algebra $\h$, where $\cI{B}0{(3)}$ is a generator, the quantum counterpart $\qI{B}0{(3)}$  is \underline{not} a generator of $\Hat{\h}$. The quantum relation under discussion implies that $\qI{B}0{(3)}$ can be generated by the generators contained in $\Hat{\h}^{0}$ and $\Hat{\h}^{1}$ due to the filtration of $\Hat{\h}$ replacing the gradation of $\h$. Actually, $\qI{B}0{(3)}$ \underline{may even be defined by this relation}, having set $g_1$ and $g_2$ equal to zero, since in this relation, $\qI{B}0{(3)}$ appears \underline{linearly} as a quantum correction. Thus instead of an expected structural similarity we are confronted with a serious dissimilarity (which, however, is not unwelcome at all). If we allow our imagination to roam, then we may even take into consideration a scenario of extreme dissimilarity between $\h$ and $\Hat{\h}$. In this scenario there exists an infinity of classical $l=$~odd, $J^P=0^+$-relations which upon application of the above mentioned quantum deformation routine one by one mutate into formulae defining some or possibly all not yet specified quantum counterparts of the classical exceptional generators $\cI{B}0{(l-2)}$ (possibly $\cI{B}0{(l-4)}, \ldots$) and of the classical generators looming in the background [cf. the comment at the end of 2. above]. In fact, this may even lead to a \underline{finitely generated} quantum algebra of string observables $\Hat{\h}$, a thoroughly welcome structure, indeed.
\\

The construction of general (non-trivial) ``physical'' string states
which i)~depend exclusively on observable, compatible data,
ii)~contain the maximal possible information for every ensemble of
strings prepared in a pure state and iii)~form a separable Hilbert
space with a positive-definite scalar product (likewise depending
exclusively on the observable, compatible data of the two states
involved) carrying a hermitean representation of the Poincar\'e
algebra and of the quantum algebra of string observables is still an
open problem. Among other characteristics, a corresponding irreducible
hermitean representation would be labelled by a fixed mass and --
apart from the trivial vacuum representation -- by a fixed (positive)
sign of $\cP^0$ as well as by a fixed number $(-1)^F$, $F=2j$
signalling the exclusive integer-valuedness, respectively, the
exclusive half-integer-valuedness of the spins $j$ appearing in its
decomposition into isotypical components w.r.t. the Poincar\'e
transformations.

The vacuum state should be invariant under translations and
rotations. It is expected to be non-degenerate implying that the
representatives of the generators of the algebra of observables
reproduce it, or even annihilate it.

There may exist physical string states with non-vanishing
energy-momentum and possibly non-vanishing spin which are very similar
to those of ordinary elementary particles in so far as they do not
have any features characteristic of strings. In other words, there may
exist physical states different from the vacuum state, on which
representatives of the generators of $\Hat{\h}$ such as the ones
contained in $\Hat{\h}^{1}$ act as annihilators. However, the
generating relations of the algebra $\Hat{\h}$ do not allow such
states unless their spin is zero or onehalf, the latter option
existing only if the (rational valued) free parameter $f$ is set equal
to $-8/5$.
This is remarkable, as this value for the free parameter $f$ was already suggested before in two seemingly different contexts~\cite{9}\cite{1}.
\\

In this communication, various aspects of the classical and the quantum algebras of string observables $\h$ and $\Hat{\h}$, respectively, as well as those of the embedding algebra $S(\fr)$ of $\h$ have been discussed. An algorithm which had been applied in the past over and over again has been justified. A simplification of the generating relations has been noted, thereby reducing the number of free parameters of the quantum algebra $\Hat{\h}$ from 3 to 1. The appearance of an ever increasing number of generators of the classical algebra has been observed as well as the reduction of the number of generators and generating relations of the quantum algebra, contained in $\bigoplus\limits_{l=-1}^6\Hat{\h}^l$. This observation sheds completely new light on the interrelation between the classical and the quantum algebra, forcing us to revise our concepts for the passage from $\h$ to $\Hat{\h}$. Briefly, we touched on the open problem of the construction of the hermitean representations of both, the quantum algebra of string observables and the Poincar\'e algebra, simultaneously and speculated about string states which are very similar to ordinary elementary particle states. \\

Let me conclude by pointing out two articles, one of which is dealing
with the Poincar\'e covariance of $\h$ and $\Hat{\h}$ \cite{10}, the
other one with the massless case for the classical algebra $\h$
\cite{11}. These articles have not been published, but they have been
put on record in the Cornell electronic archive.

\vskip2cm
\begin{center}
\noindent {\Large \bf Appendix}
\vskip0.7cm
{\large \bf A Rigorous Proof of Proposition 17 \cite{4}}
\end{center}

\vskip1cm

\noindent We consider the case mass $> 0$.

\vskip1cm

\noindent {\bf Proposition 17 \cite{4}}: {\em If division by $m$ is admitted, all invariant charges are polynomials in the standard invariants.}
\vskip1cm

\noindent It is sufficient to prove a more precise formulation of the proposition in the energy momentum rest-frame. In this reference system we consider the basis elements of the Lie algebra $\fr$, formed by the truncated tensors ({\it recte:} tensor components) at a fixed base point $(\tau,\sigma)$, $\tau \in \mathbbm R^1$, $\sigma \in S^1$: $\Rt_{\mu_1\ldots \mu_M}$ with  cyclically minimal sequences $\mu_1\ldots \mu_M$ and $\Rt_{\mu}=\cP_{\mu}=m\;\delta_{\mu,0}$. Apart from $\Rt_0$, these truncated tensors will play the role of the indeterminates in polynomials contained in the enveloping algebra of $\fr$, the symmetric algebra $S(\fr)$. By definition, each such  polynomial contains only indeterminates with tensor rank $M$ ranging from $1$ up to some $M_{\mathrm{max}}=l_{\mathrm{max}}+2<\infty$ and has a finite upper limit for the powers of the monomials involved. Thereby it also has a finite upper limit $L_{\mathrm{max}}$ of the gradation degrees of the monomials involved.

\vskip0.5cm\noindent 
For the following discussion we need the presentation of the Lie algebra $\fr$ in the form of a direct sum of the Lie algebra $G$ and the linear space $G^\prime$: $G^\prime = \bigoplus_{M=2}^\infty G^\prime_{(M)}$, with
\[
G^\prime_{(M)} = \mathcal{LH}\big\{\Rt_{0\mu_2...\mu_M}\!\! : 
\mu_2\dots\mu_M \ {\rm cycl. \ minimal}, \;
\mu_k \!\in \! \{0,1,2,3\}\; \mbox{for} \; k\! \in\! \{2,...,M\},\,
M\!\in\! \{2,3,4,...\} 
\big\}\ .
\]
Then $\fr=G\oplus G^\prime$.

\vskip0.5cm
In addition to the notion of a standard invariant $\Z_{K+1,i}$, the 
concept of an affiliated ``refined standard invariant'' $\Z^{\#}_{K+1,i}$ 
is useful:
\vskip0.3cm

\noindent {\bf \em Definition:} Let the index $K (K',...)$ take on values 
in the natural numbers and let the index $i (i',...)$ label the basis 
elements $\Rt_{\mu^{(i)}_1\ldots \mu^{(i)}_{K+1}}$ 
($\Rt_{\mu^{(i')}_1\ldots 
\mu^{(i')}_{K'+1}}, \ldots $) contained in $G_{(K+1)}$ ($G_{(K'+1)},\ldots$). The 
refined standard invariant $\Z^{\#}_{K+1,i}$ affiliated with the standard 
invariant $\Z_{K+1,i}$ is equal to the standard invariant $\Z_{K+1,i}$ 
plus a polynomial in the standard invariants $\Z_{K'+1,i'}$, where $K' < 
K$, with possibly $m$-dependent coefficients such that when all truncated 
tensors $\Rt_{\nu_1\ldots \nu_{N}}$ contained in $G^\prime$ are set to zero, only the leading linear term of $\Z_{K+1,i}$ 
survives.

Clearly, for every standard invariant there does exist a uniquely defined 
affiliated refined standard invariant, which may in some cases even 
coincide with the given standard invariant.
\vskip0.3cm

\noindent {\bf \em Remark:} Typically, the dependence of the coefficients is expressed in negative integer powers of $m$.

\vspace{0.5ex}
\noindent {\em Example:} $\Z^{\#}_{2+1,i} := (3-1)! \; \Z^{(3),\#}_{001223} = (3-1)!\; \Z^{(3)}_{001223} + \frac{1}{2m} \; \Z^{(2)}_{012} \; \Z^{(2)}_{0203}$. The only surviving leading term is $m^2\; \Rt_{1223}$, i.e. the leading term of $(3-1)!\; \Z^{(3)}_{001223}$.

\vskip0.5cm
\noindent {\bf \underline{More precise formulation of Proposition 17 [1] in the energy momentum rest-frame:}}

\noindent {\em A given polynomial $\cP_{inv}$ (over the field of the complex numbers) in $S(\fr)$: a polynomial in the basis elements of $\fr$ as indeterminates, which is invariant under arbitrary orientation-preserving reparametrizations and which does not feature a monomial consisting just of a pure non-negative power of $m$, is a polynomial (possibly with mass-dependent coefficients) in the refined standard invariants, the latter ones carrying a gradation degree $\leq l_{\mathrm{max}}$.} 

\vskip0.3cm \noindent {\bf Proof:} Because all basis elements of 
$\bigoplus_{\scriptscriptstyle{M=1}}^{\scriptscriptstyle{\infty}} 
\fr_{(M)}$ are algebraically independent of each other, we are 
allowed to consider all basis elements not equal to $\Rt_0$ as indeterminates and 
$\Rt_0= m > 0$ as being a fixed mass. According to the results of section 
III of \cite{6}, the standard invariants $\Z_{K+1,i}$, 
$i=1,\ldots,I_{K-1}$, $1 \leq I_{K-1} \leq n(4,K+1) - n(4,K)$, $K \in 
\mathbbm N$ form a complete system of algebraically independent invariant 
polynomials contained in $S(\fr)$ for fixed mass $m$, and so do the 
refined standard invariants $\Z^{\#}_{K+1,i}$.  These refined standard 
invariants are contained in $\h^{K-1}$ and, for fixed mass, they are 
themselves invariant polynomials in the basis elements of 
$\bigoplus_{\scriptscriptstyle{M=2}}^{\scriptscriptstyle{K+1}} \fr_{(M)}$ 
with characteristic 
``surviving'' linear terms (consisting each time of \underline{one} basis 
elements of the Lie algebra $G_{(K+1)}$ multiplied by the mass $m$ to some 
positive integer power). The surviving linear terms of the refined 
standard invariants vary independently of each other and independently of 
the basis elements of $G^\prime$. Because all basis 
elements of $\fr_{(K+1)}$ are algebraically independent of all basis 
elements of $\fr_{(K'+1)}$ with $K \neq K'$, for fixed mass $m$, already 
the refined standard invariants of degree $K-1$, $0 \leq K-1 \leq 
l_{\mathrm{max}}$, form a complete system of algebraically independent 
invariant polynomials with the basis elements of 
$\bigoplus_{\scriptscriptstyle{M=2}}^{\scriptscriptstyle{l_{\mathrm{max}}+2}} 
\fr_{(M)}$ as indeterminates. Thus $\cP_{inv}$ is some function $f$ of the 
refined standard invariants $\Z^{\#}_{K+1,i}$ with $1\leq K\leq 
l_{\mathrm{max}}+1$, a function still to be determined. Actually, 
$\cP_{inv}$ does not depend on all of the latter refined standard 
invariants, instead at this stage of the argumentation, a possible 
dependency on one of the latter refined standard invariants can only exist 
if its surviving linear term shows up among the indeterminates of 
$\cP_{inv}$. \vskip0.5cm

It is about time to introduce the following ideal (under tensor multiplication) $J\subset S(\fr)$:

\vspace{-\parskip}\noindent $J:=$ linear span of all monomials in the basis elements of $\fr$, which exhibit among their factors at least one basis element, contained in $G^\prime$. 

Actually, we have already called upon this ideal in the definition of $\Z^\#_{K+1,i}$ above.

Subsequently -- if desired -- all polynomials in the basis elements of $\fr$ which differ by elements of $J$ may be identified.
Thereby the refined standard invariants are replaced by their surviving linear terms, the above function $f$ of the refined standard invariants is replaced by the same function $f$, however this time with the corresponding surviving linear terms as arguments, and  $\cP_ {inv}$ is replaced by a polynomial $Q$ in the 
basis elements of $\bigoplus_{\scriptscriptstyle{l'=0}}^{\scriptscriptstyle{l_{\mathrm{max}}}} G_{(l'+2)}$, i.e. by a polynomial in the surviving linear terms of refined standard invariants of 
gradation degree $K-1$ with $0 \leq K-1 \leq l_{\mathrm{max}}$.
\vskip0.5cm

Again, by the algebraic independence of the surviving linear terms of the refined standard invariants, the function f, still to be determined, with the surviving linear terms of the refined standard invariants (affiliated to the original standard  invariants) as arguments instead of the refined standard invariants themselves, must be identical to the polynomial $Q$ with the basis elements of $\bigoplus_{\scriptscriptstyle{l=1}}^{\scriptscriptstyle{l_{\mathrm{max}}}} G_{(l+2)}$ as arguments, i.e. to $\cP_{inv} \mod J$. In particular, the function $f$ must depend on the same arguments as the polynomial $Q$, it must be a polynomial of the same polynomial degree as $Q$, and modulo ordering of its arguments, respectively of the arguments of the polynomial $Q$, it must coincide with $Q$. Thus the following equation is established: 
\[
\cP_{inv}\big(\Rt_{\mu_1\ldots\, \mu_M}: \;\mu_1,\ldots,\mu_M \; \mbox{cyclically minimal,}\; 2\leq M \leq l_{\mathrm{max}}+2\big) = Q\big(\Z^{\#}_{K+1,i}(\Rt_{\mu_1^{(i)} \ldots\, \mu^{(i)}_{K+1}},\dots)\big),
\]
where $\Rt_{\mu_1^{(i)} \ldots \, \mu^{(i)}_{K+1}}\!\! \in G_{K+1}$ is present on the r.h.s. if and only if it already appears in $\cP_{inv}\mod J$. All other $\Rt$'s on the r.h.s. form part of non-linear monomials displaying at least one factor $\Rt_{\nu_1^{(i)} \ldots\, \nu_M^{(i)}}$ which is contained in $G^\prime_{(M)}$, $2 \leq M \leq K$, $1\leq K \leq l_{\mathrm{max}}+1$.

\vskip0.5cm
\noindent Thereby the Proposition 17 of \cite{4} has been proved.

\vskip0.5cm
\noindent {\bf \em Remark:} As a spin-off, an efficient, non-recursive algorithm for finding the polynomial $Q(\Z^{\#}_{K+1,i})$ has been obtained.

\end{document}